\documentclass[aps, prx, letter, superscriptaddress, twocolumn, amsfonts, amssymb, amsmath, reprint, showkeys, nofootinbib, twoside]{revtex4-2}

\usepackage{graphicx}
\usepackage[colorlinks=false,
             pdfborder={0 0 0},
             ]{hyperref}
\usepackage{amssymb}
\usepackage{amsmath}
\usepackage{epsfig}
\usepackage{chemarr}
\usepackage{url}
\usepackage[sort&compress]{natbib}
\usepackage{dcolumn}
\usepackage{bm}
\usepackage{color}
\usepackage{braket}

\def\be{\begin{equation}}
\def\ee{\end{equation}}
\def\bmu{\begin{multline}}
\def\bea{\begin{eqnarray}}
\def\eea{\end{eqnarray}}

\begin{document}
\title{The fitness landscape of overlapping genes}

\author{Orson Kirsch}
\thanks{These authors contributed equally.}
\affiliation{Department of Physics and Astronomy, University College London, United Kingdom}

\author{Nicole Wood}
\thanks{These authors contributed equally.}
\affiliation{Department of Physics and Astronomy, University College London, United Kingdom}

\author{Steven A. Redford}
\affiliation{Department of Physics and Astronomy, University College London, United Kingdom}
\affiliation{Department of Genetics, Evolution, and Environment, University College London, United Kingdom}

\author{Kabir Husain}
\email{kabir.husain@ucl.ac.uk}
\affiliation{Department of Physics and Astronomy, University College London, United Kingdom}
\affiliation{Laboratory for Molecular Cell Biology, University College London, United Kingdom}

\begin{abstract}
Natural genomes sometimes encode two different proteins in staggered reading frames of the same DNA sequence. Despite the prevalence of these `overlapping genes' across the tree of life, it remains unknown whether arbitrary protein pairs can overlap, to what extent such overlaps are feasible, or what design principles govern them. Here, we study compatibility, frustration, and connectivity in the fitness landscape of overlapping genes. We computationally design sequences \textit{de novo} that satisfy the dual functional constraints of two distinct protein families. The joint fitness landscape, inferred via Potts models from multiple sequence alignments, reveals a fundamental trade-off between the two proteins and provides a simple criterion for when overlap is feasible. We find widespread compatibility between protein families, with one class of reading frames markedly more permissible than others. By exploring alternative genetic codes, we find that the natural genetic code is uniquely well-suited to support overlapping genes. Constructing mutational paths between sequences, we find that sequence-diverse overlapped genes can be connected via a network of near-neutral mutations. Overall, our results suggest that protein fitness landscapes are sufficiently flexible so as to accommodate the stringent, orthogonal requirements of overlapping genes.
\end{abstract}

\maketitle

\section{Introduction}

In 1957, the physicist George Gamow communicated a paper to PNAS in which the young biologist Sydney Brenner analysed the occurrence of amino acid pairs (`dipeptides') in the sparse protein sequencing data available at that time \cite{brenner1957impossibility}. On the basis of that analysis, Brenner rejected the possibility -- previously raised by Gamow himself \cite{gamow1954possible} -- that the genetic code was an overlapping code; instead arguing that non-overlapping triplets of nucleotides were the most parsimonious manner in which proteins were genetically encoded. This was soon proven to be the case \cite{crick1961general, crick1962genetic}, and the study of overlapping codons was abandoned.

However, in the early 1970s, it emerged that the virus $\phi$X174 seemingly encoded for more proteins than could be accounted for by the size of its genome. In 1977, Fred Sanger and colleagues sequenced its genome to discover that a single stretch of DNA coded for multiple proteins in staggered reading frames \cite{barrell1976overlapping,sanger1977nucleotide}. These `overlapping' genes have since been found in viruses, bacteria, and eukaryotes \cite{wright2022overlapping}. Particularly in the confined space of a viral capsid, they are thought to serve as a form of data compression \cite{chirico2010genes}. In other contexts, they have been suggested to serve as sites for the \textit{de novo} origin of genes \cite{rodin1995two, iyengar2024antisense}. More recent work has engineered overlaps in the lab between naturally non-overlapping genes, mutationally entangling otherwise unrelated functions \cite{blazejewski2019synthetic, byeon2025design, leonard2026synthetic}.

\begin{figure*}
\includegraphics[width=\linewidth]{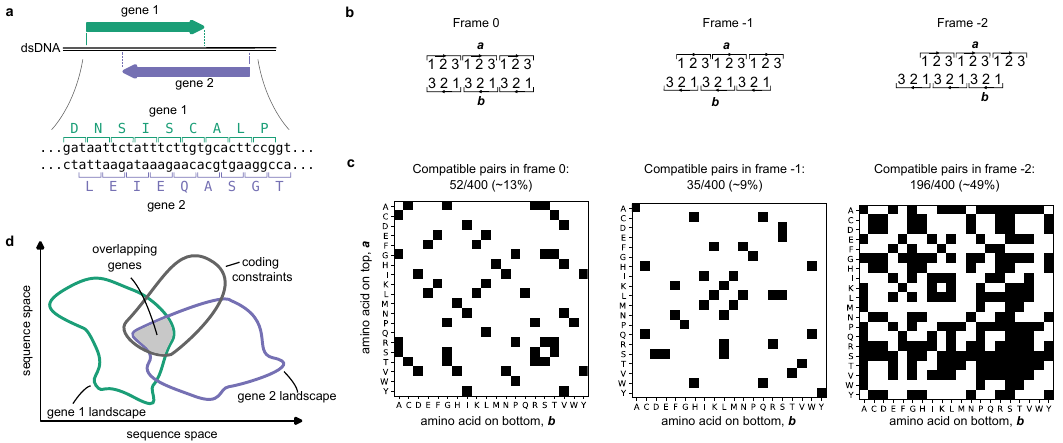}
\caption{ \label{fig:idea}
\textbf{Overlapping genes must satisfy functional as well as coding constraints.}
\textbf{(a)} Schematic of a pair of overlapping genes, in which multiple reading frames of a single coding sequence are translated into different proteins.
\textbf{(b)} Definition and nomenclature of reading frames studied in this work.
\textbf{(c)} Compatible pairs of amino acids \textit{\textbf{a}} and \textit{\textbf{b}} (as defined in (b)) that can be encoded across from each other in each of the three reading frames. 
\textbf{(d)} Diagrammatic representation of the fitness landscape of overlapping genes, which must satisfy the constraints of folding and function of each gene family, as well as the coding constraints imposed by the overlapping reading frames.
}
\end{figure*}

The prevalence of overlapping genes raises a fundamental question: can any two arbitrary genes be overlapped to an arbitrary extent -- and why or why not? On the one hand, the degeneracy of the genetic code -- with 64 codons encoding only 20 amino acids -- suggests that there may be sufficient flexibility in the mapping from DNA to protein to accommodate the constraints of two proteins simultaneously. On the other hand, the functional requirements of protein folding and activity impose stringent constraints on amino acid sequences, and it is unclear when these constraints can be jointly satisfied.

There has been a sustained interest in the evolution and design of overlapping genes \cite{krakauer2000stability,lebre2017combinatorics,opuu2017computational}. Notably, recent work has leveraged advances in computational protein design \cite{blazejewski2019synthetic, byeon2025design} to engineer synthetic overlaps between naturally non-overlapping genes. Their results argue that overlaps can be obtained with surprising ease -- with consequences both for natural evolution and synthetic biology. However, the reasons why overlaps are so easily obtained -- and their limits, if any -- have not yet been systematically explored.

Here, we take a computational approach to analysing the joint fitness landscape of overlapping genes. We leverage Potts model representations of protein fitness landscapes which are trained on multiple sequence alignments of natural protein families and have been shown to generate functional synthetic proteins in the lab \cite{morcos2011direct,cocco2018inverse, russ2020evolution}. By combining the Potts models of two protein families, we develop methods to explore their joint landscape and identify the conditions under which they can be successfully overlapped -- and when they cannot.

We find widespread compatibility between protein families, particularly in one alternative reading frame - in which the wobble base of one codon sits across from the second base of the opposite codon. Simulated `what-if' scenarios with alternative genetic codes confirm that the genetic code found in nature is particularly well-suited to support overlapping genes. We demonstrate that independently generated overlapping sequences are mutationally connected through functional intermediates. Collectively, these findings suggest that the potential for gene overlap is an intrinsic, latent feature of the standard genetic code and real protein fitness landscapes.

\section{Results}

\subsection{Overlapping imposes stringent coding constraints}

We begin by characterising the raw coding constraints imposed by overlapping.
In this work, we restrict ourselves to studying overlapping genes encoded on opposite strands of DNA. The three alternative reading frames are shown in Fig.~\ref{fig:idea}b, which we denote frames 0, -1, and -2. Due to the degeneracy of the genetic code, an amino acid sequence may be encoded on one strand by several possible choices of nucleotide sequence. As there are $61/20 \sim 3$ codons per amino acid on average, a particular amino acid sequence of length $L$ could be encoded by $3^L$ different nucleotide sequences -- with each one potentially leading to a different amino acid sequence in the alternative reading frame.

However, this flexibility is insufficient to encode any two arbitrary amino acid sequences in overlapping reading frames. One way to see this -- inspired by Brenner's dipeptide analysis of 1957 \cite{brenner1957impossibility} -- is to characterise the compatibility of amino acids across the alternative reading frames. For every amino acid \textit{\textbf{a}} on the top stand, we computationally enumerated all possible codon choices to determine whether an amino acid \textit{\textbf{b}} can be encoded opposite it in one of the alternative reading frames. 

\begin{figure*}
\includegraphics[width=\linewidth]{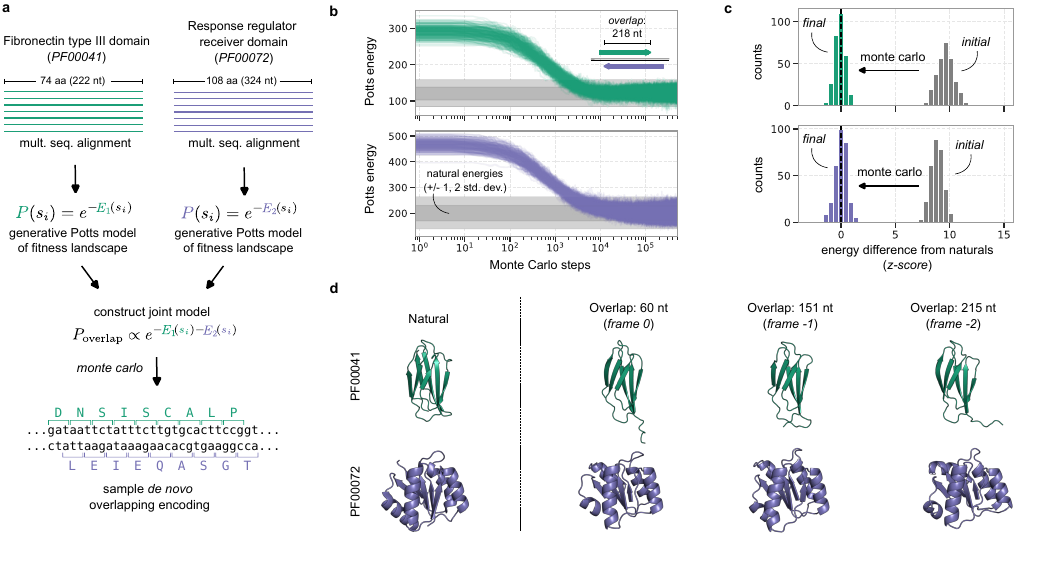}
\caption{ \label{fig:dca}
\textbf{Overlapping genes \textit{de novo} with MSA-trained generative models.}
\textbf{(a)} Sequence alignments of homologues are used to train a Potts model representation of each protein's fitness landscape, which we combine to obtain a generative model for overlapping genes.
\textbf{(b)} Potts model energies over Monte Carlo iterations for a sequence overlapping a Fibronectin type III domain (PFAM PF00041) with a two-component response regulator domain (PFAM PF00072). 300 replicate trajectories are shown, each independently initialised.
\textbf{(c)} Histogram of initial and final energies of each protein product, normalised as z-scores to the distribution of natural energies, for the different replicates in (b).
\textbf{(d)} Comparison of crystal structures of representative members of each protein family (left; PDB IDs 1TEN and 6TNE) with AlphaFold predictions for gene products from coding sequences with increasing overlap.
}
\end{figure*}

The resulting compatibility matrices are shown in Fig.~\ref{fig:idea}c. We find that of the 400 possible pairings, only 52 ($\sim 13\%$), 35 ($\sim 9\%$), and 196 ($\sim 49\%$) can be encoded in frames 0, -1, and -2, respectively. The problem worsens for longer sequences -- of the 400$\times$400 = 160,000 possible pairings of amino acid doublets, only 2704 ($\sim 1.7\%$) are compatible in frame $0$, 960 ($\sim 0.6\%$) in frame $-1$, and a comparatively larger 11,393 ($\sim 7.1\%$) in frame $-2$ (SI Fig.~S1a,b). Therefore, despite the degenerate nature of the genetic code, there are severe and non-trivial constraints on the amino acid sequences that can be encoded in overlapping reading frames.

As a consequence, if we are given two genes to overlap \textit{de novo}, we must modify the amino acid sequences of both to (a) be compatible with the coding constraints of overlapping, while (b) maintaining the folding and function of each gene. These constraints are naturally cast in terms of the fitness (function) landscapes of both genes. Coding constraints reduce the space of possible sequences, and in this reduced space a successful overlap requires sequences that are simultaneously high-fitness in both landscapes. In other words, the subset of nucleotide sequences that encode functional, properly folded versions of one protein must intersect with the corresponding subset for the other, Fig.~\ref{fig:idea}d. The feasibility of \textit{de novo} gene overlap thus depends on the existence and accessibility of such intersecting high-fitness regions.

\subsection{Overlapping genes \textit{de novo} with sequence alignment-trained Potts models}

To computationally approximate the fitness landscape of a protein family, we turned to a Potts model representation. The Potts model takes as its input the amino acid sequence $\mathbf{s}$ of a protein, where $s_i$ is the amino acid at position $i$, and outputs a statistical `energy' that acts as a computational proxy for the (negative) fitness of that sequence,
\begin{equation} \label{eq:pottsenergy}
    E(\mathbf{s}) = \sum_{i} h_i(s_i) + \sum_{ij} J_{ij}(s_i, s_j).
\end{equation}
The `fields' $h_i$ and `couplings' $J_{ij}$ are learned for each protein family from a multiple sequence alignment of natural sequences from that family. In particular, the learning procedure (known as `direct coupling analysis', or DCA \cite{morcos2011direct, figliuzzi2018pairwise}) seeks to determine $h_i$ and $J_{ij}$ in such a way that sequences sampled from the Boltzmann distribution,
\begin{equation}\label{eq:boltzmann}
    P(\mathbf{s}) \propto \exp(-E(\mathbf{s})/T),
\end{equation}
reproduce the single-site and two-site statistics of the sequence alignment. Here, in analogy with equilibrium statistical mechanics, $T$ is a sampling `temperature' -- at low $T$, only the lowest-energy (and presumably highest-fitness) sequences are sampled, while at high $T$, the constraints on the protein are relaxed and high-energy (low-fitness) sequences are sampled. The temperature $T$ set is to $T=1$ for training, but is often lowered during sampling such that the distribution of sampled energies $E(\mathbf{s})$ matches the distribution of energies of natural sequences (i.e. those from the multiple sequence alignment) \cite{fields2025understanding}. Widely studied in recent years \cite{figliuzzi2018pairwise,ngampruetikorn2022inferring,di2024emergent}, sequences sampled from Potts models have been experimentally found to fold and function as well as their natural counterparts \cite{russ2020evolution}.

We take the learned Potts model energy function $E(\mathbf{s})$ of a protein family as (the negative of) an approximation to its true fitness landscape. While other methods have been proposed -- such as Large Language Models or other deep learning techniques \cite{khakzad2023new, lian2024deep,kantroo2025pseudo,hayes2025simulating} -- Potts models have the advantage of simplicity and computational tractability. Nonetheless, as we describe further in the Discussion, its use as a proxy for fitness is fraught with unknowns and requires, in the final analysis, experimental verification and potential model refinement. We leave this to future work.

We sought to repurpose Potts models of protein fitness landscapes as a method to generate overlapping genes. To do so, we took advantage of the fact that Potts models operate on the amino acid sequence $\mathbf{s}$ of a protein, while the constraints on an overlapping gene come from its underlying DNA coding sequence, $\mathbf{d}$. We wrote a modified Monte Carlo algorithm that operates directly on the DNA coding sequence. The degree of overlap is specified at the outset, such that the reading frames of each gene are fixed for every Monte Carlo run. Mutations are made directly at the nucleotide level, and are accepted or rejected based on the change in the \textit{joint energy},
\begin{equation}\label{eq:joint}
    \mathcal{H} = \frac{E_1(\mathbf{s}_1(\mathbf{d}))}{T_1} + \frac{E_2(\mathbf{s}_2(\mathbf{d}))}{T_2},
\end{equation}
\noindent where $\mathbf{s}_1(\mathbf{d})$, $\mathbf{s}_2(\mathbf{d})$ denote the amino acid sequences of genes encoded in the two reading frames of $\mathbf{d}$. $T_1$ and $T_2$ are independent temperatures controlling the stringency of selection for each protein, setting the relative weighting of each individual protein's constraints in the joint fitness landscape. We use a standard Metropolis acceptance scheme: mutations are accepted if they decrease the joint energy, $\Delta\mathcal{H} < 0$, or accepted with probability $e^{-\Delta\mathcal{H}}$ if they increase the joint energy, $\Delta \mathcal{H} > 0$. Nucleotide mutations that lead to a premature stop-codon in either reading frame are rejected outright.

\begin{figure*}
\includegraphics[width=\linewidth]{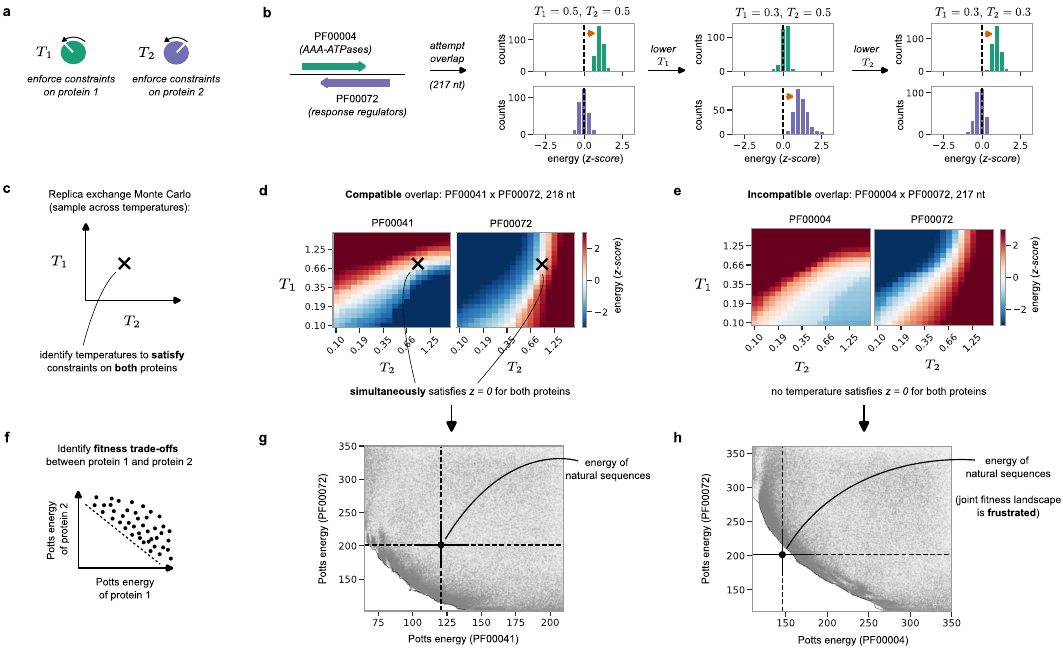}
\caption{ \label{fig:remc}
\textbf{Replica exchange Monte Carlo maps the joint fitness landscape of overlapping genes.}
\textbf{(a)} Decreasing temperature $T_1$ ($T_2$) acts as an increased selection pressure that favours the constraints of protein family 1 (2) in the sampled sequence.
\textbf{(b)} Histograms of sampled energies, normalised as z-scores, for sequences sampled at indicated temperatures. For this overlap (PF00004 $\times$ PF00072 at 217 nt), decreasing the temperature of one protein increases the other energy of the other (indicated by an orange arrowhead).
\textbf{(c)} We implement a replica exchange Monte Carlo method to scan over all pairs of temperatures.
\textbf{(d, e)} Heatmaps of z-scores for indicated protein pairs. Marked cross on left indicates the point at which both proteins simultaneously achieve the energy of the naturals (i.e. a z-score of 0).
\textbf{(f)} Schematic of analysis, in which the energies of sequences sampled at all temperatures are plotted to identify a trade-off between satisfying the constraints of each protein family. 
\textbf{(g, h)} Trade-off analysis for indicated protein pairs. Grey points are samples from the joint fitness landscape. Dashed lines and black marker indicate the natural energies, with solid black arms denoting standard deviations of the naturals. In (h), the natural energies lie outside the region of achievable energies -- indicating that the landscape is `frustrated' and an overlap is not feasible.
}
\end{figure*}

As a proof of principle, we first attempted to overlap a Fibronectin type III domain (PFAM family PF00041) with a two-component response regulator receiver domain (PF00072). We chose these protein families because they are well-characterized, have deep multiple sequence alignments, and represent distinct structural folds and functional classes. The Fibronectin type III domain adopts an immunoglobulin-like beta-sandwich fold and is found in a wide variety of extracellular proteins, while the response regulator receiver domain is an alpha/beta protein central to bacterial signal transduction. Members of the former are found primarily in eukaryotes while members of the latter are predominantly bacterial.

After downloading and preprocessing the sequence alignments of each protein family, we used the software package adabmDCA \cite{rosset2012adabmdca} to infer the parameters of the DCA model for each protein (see Methods for details). We then implemented our modified Monte Carlo procedure, setting as a goal that both gene products achieve an energy, Eq.~\ref{eq:pottsenergy}, comparable to the distribution of energies of their natural homologues (i.e. sequences in their respective MSAs). We quantify this via a z-score, defined for each gene product:
\begin{equation}\label{eq:zscore}
    z = \frac{\langle E \rangle - \mu}{\sigma},
\end{equation}
\noindent where $\langle E \rangle$ is the average energy of the sampled sequences, and $\mu$, $\sigma$ are the mean and standard deviation of the natural energies. The goal is to achieve $z \approx 0$ for both gene products.

In Fig.~\ref{fig:dca}b, we show Monte Carlo trajectories for an overlap of 218 nucleotides -- a level of overlap at which the smaller protein (PF00041) is almost entirely nested within the coding region of the larger protein (PF00072). We initialised our Monte Carlo simulations with random DNA sequences and tracked the Potts model energies of both gene products over successive iterations. The energies of both proteins decreased over the course of the simulation, indicating that the algorithm successfully navigates the joint fitness landscape to find sequences that satisfy the constraints of both protein families simultaneously. The energies plateau after approximately $10^5$ Monte Carlo steps. We find that the ensemble, sampled here at $T_1 = 0.7$ and $T_2 = 0.88$, results in gene products with a distribution of energies that converges to those of the natural proteins, Fig.~\ref{fig:dca}c, indicating a successful overlap.

We repeated the procedure at overlaps of 60, 151, and 215 nucleotides -- corresponding to overlaps in alternate frames 0, -1, and -2, respectively. At each, we found that we were able to retrieve sequences with energy comparable to the naturals. Taking one sequence from each level of overlap, we used AlphaFold \cite{abramson2024accurate} to predict the folded structure of each gene product. These are shown alongside representative crystal structures of their natural counterparts in Fig.~\ref{fig:dca}d. Both the beta-sandwich fold of the Fibronectin domain and the alpha/beta fold of the response regulator were faithfully recapitulated, providing an orthogonal prediction that the sequences found by our method are likely to encode functional proteins.

\subsection{Replica exchange sampling of the joint fitness landscape identifies signatures of compatible and incompatible overlaps}

\begin{figure*}
\includegraphics[width=\linewidth]{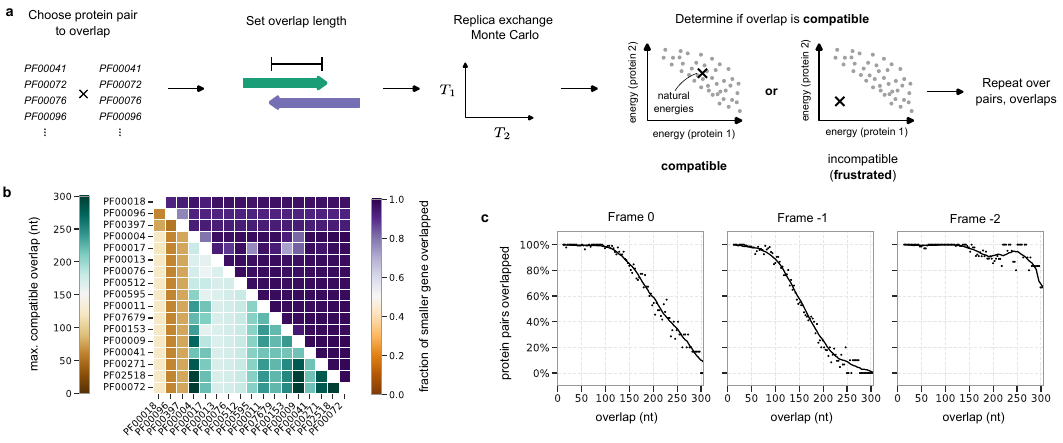}
\caption{ \label{fig:minus2}
\textbf{A systematic overlap of protein families finds widespread compatibility in the -2 reading frame.}
\textbf{(a)} Schematic workflow, in which we survey the compatibility of overlap between all pairs of 17 protein families.
\textbf{(b)} Heat map showing the maximum compatible overlap length for each pair as (lower left) nucleotides, and (upper right) a fraction of the smaller gene.
\textbf{(c)} Protein-family pairs successfully overlapped, as a function of overlap length and split by reading frame. Solid line is a smoothened curve, shown as a guide to the eye.
}
\end{figure*}

Buoyed by this success, we attempted to overlap other pairs of proteins. However, we found that this was not always so straightforward. In Fig.~\ref{fig:remc}a,b we demonstrate this with PF00004 (i.e., the family of AAA-ATPases) and PF00072 (two-component response regulators) at an overlap of 217 nucleotides. For this pair of proteins, lowering $T_1$ to improve the energy of protein 1 inevitably pushed up the energy of protein 2, and vice versa. It appeared that no choice of temperatures $(T_1, T_2)$ could simultaneously satisfy both constraints -- suggesting a trade-off that could not be resolved. Our failure motivated a systematic approach to map the joint fitness landscape.

To systematically determine if, and when, overlapping is feasible, we implemented an analysis based on replica exchange Monte Carlo (REMC), also known as parallel tempering \cite{swendsen1986replica, marinari1992simulated}. REMC simultaneously equilibrates a population (`replicas') of sequences, with each equilibrating at a different pair of temperatures $(T_1,T_2)$. Regular `swaps' of sequences between replicas speeds up convergence, allowing all temperatures to be rapidly and simultaneously equilibrated. 

We use replica exchange sampling to ask if there exists any temperature pair at which an overlap is successful. As before, we define a successful overlap as one in which both gene products have an energy comparable to the natural distribution of homologues for each. 

Fig.~\ref{fig:remc}c,d,e shows the results of replica exchange equilibration for both protein pairs tried previously. The z-score landscape for PF00041 $\times$ PF00072 at an overlap of 218 nucleotides (as in Fig.~\ref{fig:dca}) reveals the expected structure: $z_1$ decreases with decreasing $T_1$ (stronger selection on protein 1), while $z_2$ decreases with decreasing $T_2$. As anticipated, we find a pair of temperatures that simultaneously achieves $z=0$ for both proteins (black cross, Fig.~\ref{fig:remc}d). In contrast, the z-score landscape for PF00004 $\times$ PF00072 with a 217 nucleotide overlap confirms that there exists no pair of temperatures at which both z-scores are simultaneously $\approx 0$ -- that is, the constraints of the two families are in fundamental tension at this degree of overlap.

We combine data from across all temperature pairs and plot the joint distribution of $(E_1, E_2)$ values, sampling widely across the energy landscape (Fig.~\ref{fig:remc}f). In both PF00041 $\times$ PF00072 and PF00004 $\times$ PF00072, we observe a trade-off, or `Pareto', front that demarcates the energies that can be achieved by sequences in the joint landscape. Crucially, for PF00041 $\times$ PF00072 at an overlap of 218 nucleotides, the natural mean energy $(\mu_1, \mu_2)$ lies well within the accessible cloud of sampled energies, confirming that overlapping sequences matching the energetics of both natural protein families are readily achievable, Fig.~\ref{fig:remc}g. However, for PF00004 $\times$ PF00072 at an overlap of 217 nucleotides, the natural mean energies falls outside the trade-off front, Fig.~\ref{fig:remc}h, indicating that no sequence in the landscape achieves energies comparable to that of both natural proteins. 

We conclude that an attempt to overlap genes may fail due to `frustration' in the joint landscape. In physical systems that exhibit disorder, `frustration' refers to the inability of a system to minimize the energy of all its interactions simultaneously \cite{toulouse1987theory,mezard2009information}. In this case, we find an analogous situation -- satisfying the constraints of both protein families are simply not simultaneously achievable by any sequence in the landscape. 

\subsection{Widespread compatibility between protein families in the -2 reading frame}

\begin{figure*}
\includegraphics[width=\linewidth]{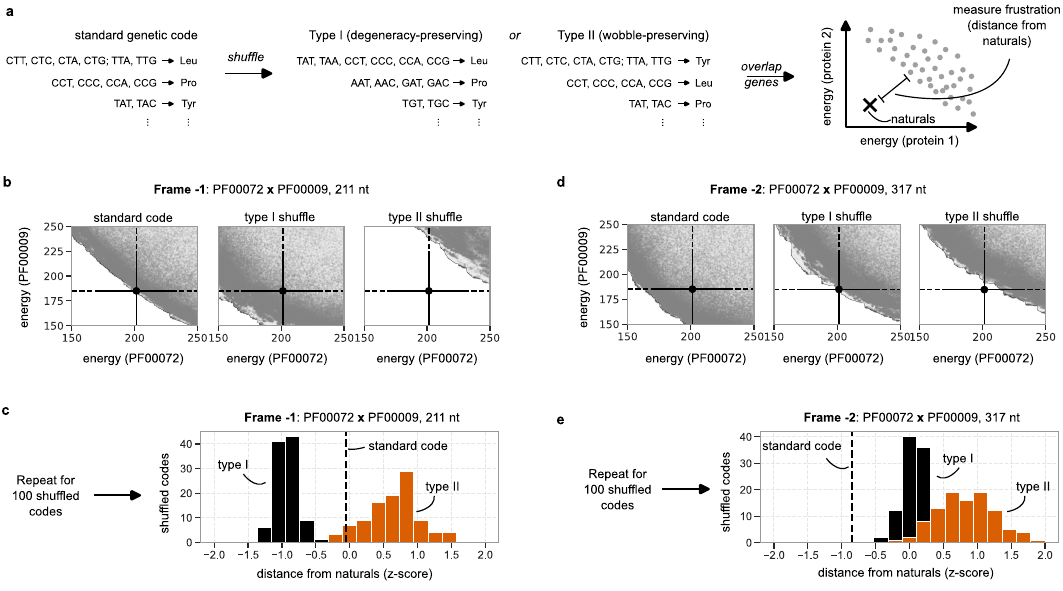}
\caption{ \label{fig:altcode}
\textbf{The standard genetic code is more permissive to overlapping than randomised codes.}
\textbf{(a)} Schematic of analysis, in which we shuffle the genetic code to produce two classes of randomised codes: type I, which preserves the degeneracy of the standard code (i.e. number of codons for each amino acid), and type II, which preserves the synonymous structure of the standard code (i.e., synonymous mutations in the standard curve remain synonymous). For each shuffled code, we repeat the trade-off analysis of Fig.~4 and measure the distance of the natural energies from the trade-off front.  
\textbf{(b, d)} Trade-offs for indicated overlapping pairs, computed for the standard code as well as one representative example each of the type I and type II shuffled codes.
\textbf{(c, e)} Histograms of distance of the natural energies (in units of the z-score) from the trade-off fronts, computed over 100 type I and type II shuffled codes. Positive values indicate a frustrated overlap (i.e., natural energies are outside the trade-off front), while negative values indicate a feasible overlap (i.e., natural energies are inside the trade-off front). Dashed line is value under the standard code.
}
\end{figure*}

The trade-off analysis provides a diagnostic tool with which to determine the compatibility of overlapping two proteins: overlap is feasible when the natural mean energies lie within the trade-off front. We used this criterion to systematically survey the compatibility of overlap between all pairs of seventeen protein families, listed in Table S1. We varied the amount of overlap from a modest 12 nucleotides to almost the entirety of the shorter gene. For each of the 136 unique pairwise combinations, at each level of overlap, we equilibrated replica ensembles, constructed trade-off fronts, and assessed whether the target natural energies lay within or outside the achievable regime, Fig.~\ref{fig:minus2}a.

In Fig.~\ref{fig:minus2}b we report the largest overlap length at which we could overlap each pair of protein families. In all cases, we found that we could successfully overlap genes at lengths approaching the size of the smaller gene -- that is, the maximal possible overlap.

How does this result reconcile with the failure to overlap in Fig.~\ref{fig:remc}e,h? We realised that a protein pair that is readily entangled at large overlaps might be incompatible at a \textit{smaller} overlap -- that is, that the constraints on the sequence are not simply monotonic in the length of shared nucleotides. Instead, we found that the strongest determinant of overlap success was the choice of reading frame. The strength of this effect varied across pairs of proteins. For instance, PF00041 $\times$ PF00072 overlapped well in all reading frames at all overlap lengths. In contrast, PF00041 $\times$ PF00271 showed large differences between reading frames -- successfully overlapping up-to $65\%$, $42\%$, and $94\%$ of the smaller gene ($\sim 200$ nucleotides) in reading frames $0$, $-1$, and $-2$, respectively (see Fig.~S1c,d).

In Fig.~\ref{fig:minus2}c we plot the fraction of compatible protein pairs as a function of overlap amount, split by each of the three reading frames. 
In each, success rate decreased with increasing overlap, as expected. However, the rate of decline differed between reading frames. 
Frames $0$ and $-1$ were statistically unlikely to support overlaps beyond 200 and 150 nucleotides, respectively, with success rates falling $< 50\%$ at larger overlaps. In contrast, overlaps in frame $-2$ remain broadly successful up-to 300 nucleotides of overlap -- at the upper end of the overlaps we attempted in this study. We conclude that the $-2$ reading frame is remarkably permissive to overlapping genes, and drives the widespread compatibility found across protein family pairs.

Why is the $-2$ frame most amenable to overlapping? One possible reason is that, in this configuration, the third (`wobble') position of each codon on one strand sits opposite the second position of the codon on the other strand (Fig.~\ref{fig:idea}b). The wobble position is well known to be the least constrained: mutations there are far more likely to be synonymous and leave the amino acid unchanged \cite{crick1966codon,freeland1998genetic}. In contrast, the second codon position is the most constrained -- no single-nucleotide mutation at the second position of any codon is synonymous. The $-2$ frame therefore pairs the most flexible position of one codon with the most constrained position of the other, maximising flexibility.

By the same logic, the $-1$ frame -- in which the two wobble positions are opposite each other -- is the least permissive: flexibility on one strand is `wasted' opposite flexibility on the other, leaving little room to satisfy the more constrained positions of either codon. This asymmetry, consistent with Fig.~\ref{fig:minus2}c, is directly reflected in the compatibility analysis of Fig.~\ref{fig:idea}c, which showed that the $-2$ frame admits roughly five times more compatible amino acid pairs (and ten times more doublet pairs, Fig.~S1b) than frame $-1$. This suggests that the structure of the genetic code itself -- the degeneracy of amino acid coding and the synonymous relationships between different codons -- could dictate the overlappability of genes in different reading frames.

\subsection{Alternate genetic codes emphasise the unique flexibility of the standard genetic code}

\begin{figure*}
\includegraphics[width=\linewidth]{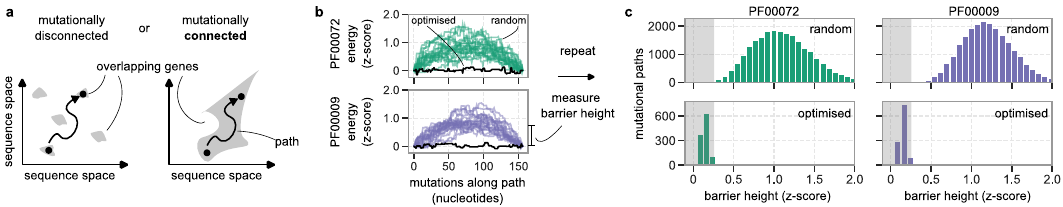}
\caption{ \label{fig:connectivity}
\textbf{Overlapping genes can be mutationally connected.}
\textbf{(a)} Schematic of connected and disconnected regions of sequence space. When sequence space is connected, a mutational path of can be constructed between two sequences along which every intermediate is functional.
\textbf{(b)} Potts energy (top: protein 1 in green, bottom: protein 2 in purple) along mutational paths between sequences encoding PF00072 $\times$ PF00009 with a 371 nucleotide overlap. Solid black line is an optimised path, coloured lines show random paths.
\textbf{(c)} Distribution of maximum z-scores along (top) random and (bottom) optimised mutational paths for PF00072 (left) and PF00009 (right). Shaded region indicates the viability threshold,  $\vert z \vert < 0.25$.
}
\end{figure*}

To test the idea that the code itself could influence the ease of finding an overlapped solution, we performed a series of `what-if' simulations with alternative genetic codes. In the first scheme (denoted type I in Fig.~\ref{fig:altcode}a) we shuffled the genetic code by randomly reassigning which amino acid each codon encodes, while preserving the overall degeneracy structure -- that is, maintaining the same distribution of codon counts per amino acid (e.g., six codons for leucine, one for methionine). This randomisation disrupts the mutational relationships between synonymous codons and, in particular, abolishes the special role of the wobble position.

We generated 100 shuffled type I codes and, for each, sampled the overlapping gene landscape for a representative Frame -1 overlap (PF00072 $\times$ PF00009, 211 nucleotides) and a representative Frame -2 overlap (PF00072 $\times$ PF00009, 211 nucleotides). We find that the overlap in frame -1 is easier with the shuffled codes, while the overlap overlap in Frame -2 is harder -- quantified by the distance of the natural energies from the sampled trade-off fronts, Fig.~\ref{fig:altcode}b-e. This result is consistent with the idea that the flexibility of the wobble position is what makes overlaps difficult in frame -1, and those in frame -2 easy.

We next employed a second randomisation scheme, in which we reassigned amino acid identities \textit{en bloc}: all codons encoding one amino acid (e.g., leucine) were reassigned to encode a different amino acid (e.g., methionine), and vice versa. This procedure, denoted type II in Fig.~\ref{fig:altcode}a, alters the degeneracy of the code -- the number of codons per amino acid -- but preserves the synonymous relationships between codons. In particular, mutations at the wobble position remain more likely to be synonymous than mutations at the first or second position. 

The results of this second set of simulations are shown (for the same overlap configurations as the type I codes) in Fig.~\ref{fig:altcode}b-e. Strikingly, overlaps using the type II codes were significantly worse in both frames -1 and -2. This implies that the structure of synonymous mutations, and the flexibility of the wobble position, are not by themselves sufficient to account for the performance of the standard genetic code. Thus it appears that, in the space of possible genetic codes that preserve either the degeneracy structure or the pattern of synonymous mutations, the natural code is exceptionally well-suited to permitting overlapping genes. 

\subsection{Overlapping genes are mutationally connected}

Finally, we asked whether overlapping genes form a connected network in sequence space. As discussed by Maynard Smith in his conception of `protein space', evolution by natural selection requires that functional sequences be connected by paths of viable intermediates that each maintain biological function \cite{maynard1970natural}, Fig.~\ref{fig:connectivity}a. Similar ideas have been fruitfully explored in RNA fitness landscapes \cite{fontana1998continuity, schuster1999chance}. We sought to determine whether overlapping genes occupy similarly connected regions of sequence space.

We generated $50$ independent sequences encoding for PF00072 $\times$ PF00009 at an overlap of 317 nucleotides, defining a functional sequence by a stringent threshold of $\vert z \vert < 0.25$ for the energies of each gene product. On average, each sequence differed from the others by $\sim160$ nucleotides, with each gene product differing at $\sim 50$ amino acids per protein, Fig.~S1e. Then, for each pair of independently generated sequences, we sought a mutational ordering -- of the $N$ nucleotide differences between them -- along which both gene products maintain Potts energies within this threshold. This is a combinatorial problem ($N!$ possible orderings); we employed a genetic algorithm \cite{katoch2021review} with crossover and swap mutations, with `fitness' defined as the maximum z-score (`barrier height') encountered along the path, see Methods for details.

Results from this analysis, for all 1225 pairwise-paths between 50 sequences, are shown in Fig.~\ref{fig:connectivity}b,c. We consistently found mutational paths with z-scores within the viability threshold, indicating that independently generated overlapping sequences are mutationally connected through functional intermediates. We conclude that, within the Potts model, one overlapping solution can be transformed into another without passing through a non-functional intermediate, suggesting that overlapping genes can occupy contiguous, navigable regions of sequence space.

\section{Discussion}

We have demonstrated that overlapping genes can be computationally designed \textit{de novo} with Potts model representations of protein fitness landscapes. Re-purposing replica exchange Monte Carlo sampling from statistical physics, we show how to determine whether an overlap is feasible or not. Our survey finds that proteins are broadly compatible, with up-to 300 nucleotide overlaps in the appropriate choice of reading frame. Our results further indicate that this is not simply a property of protein fitness landscapes at the amino acid level, but that the particular architecture of the genetic code plays a key role. Finally, we find that independently generated overlapping sequences are mutationally connected within our model, arguing that overlapping genes could form navigable networks in sequence space that evolution can traverse.

Our results come with several caveats. Most notably, while Potts models have been validated experimentally for single protein families \cite{russ2020evolution}, their accuracy as fitness predictors for sequences that deviate substantially from the training distribution -- as our overlapping sequences necessarily do -- remains uncertain. In particular, the z-score metric we employ provides a useful heuristic for sequence quality, but whether or not it is a sufficient metric requires further experimental validation. This is particularly true of mutational paths between overlapping proteins, whose validity requires that the Potts model faithfully recapitulate the detailed topology of the real fitness landscape \cite{mauri2023transition}. Given that Potts models do not account for any epistasis beyond second-order, this is unlikely to be true -- but how faithful an approximation it is remains to be tested.

In our analysis of trade-offs, compatibility, and frustration in overlapping genes, Figs.~\ref{fig:remc} and \ref{fig:minus2}, we have relied on a systematic exploration of the joint energy landscape using replica exchange Monte Carlo. While this approach is robust, it is not exhaustive -- and may miss rare states, particularly if the underlying Potts models is `glassy' \cite{mezard2009information}. However, while we cannot guarantee an exhaustive sampling of every possible state, the comparisons between different reference frames and protein pairs remain valid, as they are statements on the relative ease with which overlapping sequences can be found in the landscape.

We build on recent work \cite{blazejewski2019synthetic, byeon2025design} in several ways. First, we not only determine when overlaps are feasible, but also discuss situations in which they are not -- suggesting that, despite being easier than expected, overlapping is not necessarily universally straightforward. Expectedly, we identify overlap length as a key parameter -- with longer overlaps typically being harder to achieve. Less expectedly, the three possible reading frames on the opposite strand differ in their capacities to support overlaps -- with the -2 frame easily accommodating overlaps up-to 300 nucleotides.

Second, our results suggest that, surprisingly, overlapping genes might be mutationally connected. While it has been generally appreciated that overlapping reduces the mutational tolerance of a gene (as mutations in the overlapping region affect both genes at once) \cite{blazejewski2019synthetic, leonard2026synthetic}, our study suggests that this loss of tolerance does not compromise the mutational connectivity of sequence space. That is, while it is certainly true that overlapping reduces the number of accessible neutral mutations, there appear to be a sufficient number so as to connect different parts of sequence space. This kind of topological connectivity, often associated with evolvability \cite{payne2019causes}, argues against overlapping genes being so constrained so as to be `evolutionary dead-ends'.

Third, our work highlights the unique flexibility of the standard genetic code in permitting long overlaps between genes. It was notable that none of the shuffled codes we generated were able to match the performance of the standard genetic code. Our study leaves unanswered the question of what precise aspect of the standard genetic code makes it so flexible -- as neither codon degeneracy nor their synonymous structure were sufficient on their own. It could be a combination of these factors, or more subtle factors involving the chemical relatedness of different amino acids and general error-correction properties of the code \cite{woese1966fundamental,koonin2009origin,freeland1998genetic}.

In conclusion, we note that the problem we have analysed is a biological variant of a `satisfiability' problem \cite{mezard2009information}. These problems are known to be difficult to be solve. In that respect, it is notable that the structure of (data-derived) fitness landscapes -- and that of the genetic code -- are such that they so readily admit solutions.

\section{Methods}

All code necessary to reproduce our results is available at \href{https://github.com/kabirhusain/fitnesslandscape_olg}{github.com/kabirhusain/fitnesslandscape\_olg}.

\subsection{Alignment preprocessing and bmDCA training}

Multiple sequence alignments (MSAs) for protein families were obtained from the Pfam database (EBI InterPro) in Stockholm format. Alignments were trimmed by removing columns with more than 20\% gaps, followed by removal of sequences with more than 20\% gaps. Boltzmann machine direct coupling analysis (bmDCA) was performed on each trimmed MSA using the adabmDCA 2.0 software package \cite{rosset2012adabmdca} with default parameters. After training, the DCA energy of each natural sequence in the trimmed MSA was computed, and the resulting distribution was summarised by its mean and standard deviation to establish a family-specific baseline against which designed sequences could be evaluated.

\subsection{Initial conditions and fixed-temperature Monte Carlo}

A random DNA sequence was generated as the initial condition, subject to the constraints that it encodes two proteins in overlapping reading frames, each ORF terminates with a stop codon, and neither frame contains any internal stop codon. Sequence optimisation proceeded via single-temperature Metropolis Monte Carlo: at each step, a single nucleotide was chosen uniformly at random and mutated to a different randomly selected nucleotide; proposed mutations introducing an internal stop codon in either reading frame were rejected outright, while accepted moves followed the standard Metropolis criterion applied to the combined reduced energy Eq.~\ref{eq:joint}. Typical runs comprised $\sim 10^6$ Monte Carlo steps, with independent trials initiated from distinct random starting sequences.

\subsection{Replica exchange Monte Carlo}

To sample the joint sequence space at multiple temperature combinations simultaneously, replica exchange Monte Carlo (REMC) was implemented over a 2D temperature grid. Replicas was constructed with log-spaced temperatures, as we found we had more uniform swap acceptance rates compared with a linear temperature scale. Each replica performed standard Metropolis Monte Carlo at its assigned temperature pair using the mutation scheme described in Section 2. Every $N_{\text{swap}}$ local Monte Carlo steps, swap attempts were made between neighbouring replicas in a checkerboard pattern, alternating between horizontal exchanges (adjacent $T_1$, fixed $T_2$) and vertical exchanges (fixed $T_1$, adjacent $T_2$). 

A proposed swap between replica $i$ with per-gene temperatures $T_{1i}, T_{2i}$ and energies $E_{1i}, E_{2i}$, and replica $j$ with $T_{1j}, T_{2j}$ and energies $E_{1j}, E_{2j}$, was accepted with probability 

\begin{widetext}
$$
p_{\rm swap} = \min\!\left(1,\, \exp\!\left[\left(\frac{1}{T_{1i}} -         
  \frac{1}{T_{1,j}}\right)(E_{1i} - E_{1j}) + \left(\frac{1}{T_{2i}} -             
  \frac{1}{T_{2j}}\right)(E_{2i} - E_{2j})\right]\right).   
$$
\end{widetext}
By construction, this acceptance probability satisfies detailed balance in the product ensemble, guaranteeing convergence to the Boltzmann distribution Eq.~\ref{eq:boltzmann} for each replica; see \cite{swendsen1986replica, marinari1992simulated} for details.

The first 20\% of rounds were discarded as burn-in, and samples were collected at thinning intervals of 1,000 steps to reduce autocorrelation. From the resulting ensemble, the Pareto front of $(E_1, E_2)$ samples was identified as the non-dominated set -- i.e., points for which no other sample simultaneously achieved lower $E_1$ and lower $E_2$ -- and energies were converted to z-scores using the natural-sequence energy statistics of each family. The signed distance from the natural-sequence reference point ($z_1$ = 0, $z_2$ = 0) to the closest point on the Pareto front was then computed to determine whether the natural energies lie in or out of the Pareto front.

\subsection{Compatibility across protein families}

To assess the feasibility of designing overlapping genes across diverse protein families, the replica exchange protocol described above was applied systematically to all 136 pairwise combinations of 17 Pfam protein families, listed in Table ~S1. For each pair, the overlap length was scanned from 12 nucleotides up-to a maximum set by the shorter protein in the pair (we chose $L - 6$, where $L$ is the length of the shorter protein in nucleotides), and at each combination the resulting Pareto front in Potts energy space was identified and transformed to z-score coordinates using the marginal energy distributions of each family. Two summary statistics were recorded per (pair, overlap) condition: a boolean indicator of whether the natural mean energies fall within the Pareto front, and a signed z-score distance from that front, with negative values denoting that the natural mean lies inside the front.

\subsection{Shuffled genetic codes}

Type I shuffled codes were constructed to retain the degeneracy structure of the standard code -- each amino acid retained its original codon count -- but the assignment of sense codons to amino acids was randomised by drawing a uniform random permutation over the 61 sense codons, leaving the three stop codons (TAA, TAG, TGA) unchanged. 

In the type II shuffled codes, the synonymous codon groups were preserved exactly, but the identities of the 20 amino acids were randomly relabelled by a permutation of amino acid names, again leaving stop codons unchanged. 

\subsection{Genetic algorithm for mutational paths}

To assess whether one overlapping-gene sequence could be transformed into another through a series of individually tolerable point mutations, a genetic algorithm \cite{katoch2021review} was employed to search for optimal mutational paths. Given a start and end sequence differing at $n$ nucleotide positions (their Hamming distance), the task was to find an ordering of those $n$ single-nucleotide substitutions that minimised the maximum energetic deviation from natural sequences encountered along the path. At each intermediate sequence, the deviation was quantified as $\vert z_1 \vert + \vert z_2 \vert$, where $z_i$ is the z-score of the DCA energy for protein $i$; the fitness of a path was defined as the maximum deviation over all mutational steps. Any path traversing an internal stop codon was assigned infinite fitness. 

The genetic algorithm maintained a population of 100 `individuals', each representing a permutation of the $n$ mutation indices (i.e., an ordered sequence of mutations along the path). The initial population was generated by random sampling with rejection of any permutation whose induced mutational path contained an internal stop codon. At each generation, the `fitness' of each path was evaluated. We applied a tournament selection with a tournament size of three. Variation was introduced each generation via recombination-like `OX1 crossovers' \cite{davis1985applying}, and each path was subject to swap `mutations' that interchanged the order of steps along the path. At each generation, the top 10\% were copied over unchanged. The algorithm ran for a fixed 300 generations, after which the path with the lowest z-score value was taken as the optimal mutational path.

\section{Acknowledgments}

We are grateful to Maryn Carlson, Emily Hinds, Alex Fedorec, Marianne Bauer, Wenying Shou, and Jan Kocka for insightful and enjoyable discussions. SAR is supported by EMBO postdoctoral fellowship number ALTF 491-2024. The authors acknowledge the use of the UCL Myriad High Performance Computing Facility (Myriad@UCL) and the DIAS cluster in the Department of Physics and Astronomy, and associated support services, in the completion of this work.

\bibliography{overlapping}

\end{document}


\title{Supplemental Material\\The fitness landscape of overlapping genes}
\author{Orson Kirsch$^{1,\ast}$, Nicole Wood$^{1,\ast}$, Steven A. Redford$^{1,2}$, Kabir Husain$^{1,3}$}
\affiliation{
${}^1$Department of Physics and Astronomy,
${}^2$Department of Genetics, Evolution, and Environment,
and ${}^3$Laboratory for Molecular Cell Biology, University College London, United Kingdom.
${}^{\ast}$These authors contributed equally.}

\maketitle

\renewcommand{\thefigure}{S\arabic{figure}}
\renewcommand{\thetable}{S\arabic{table}}

\onecolumngrid

\begin{figure*}
\includegraphics[width=\linewidth]{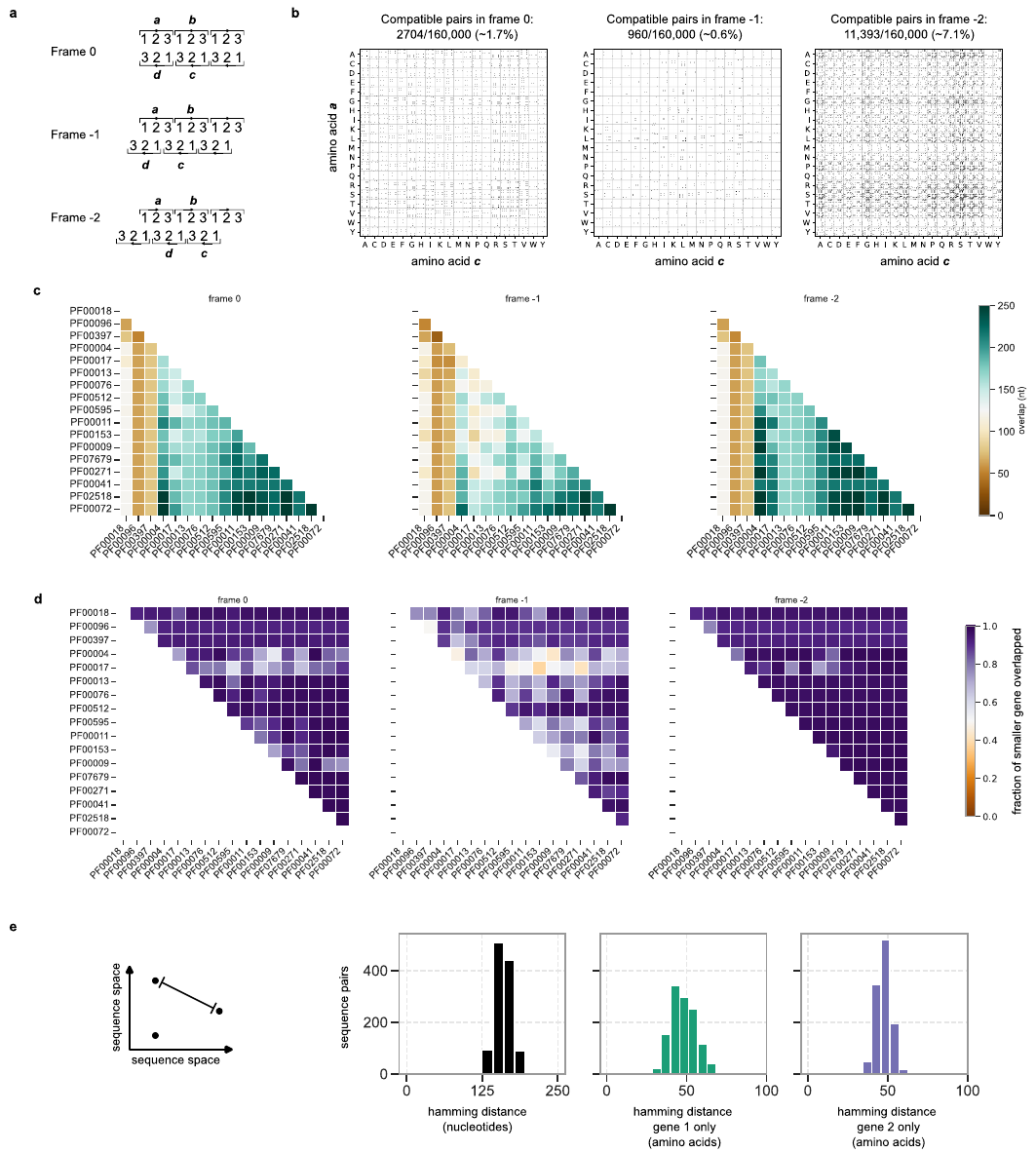}
\caption{ \label{fig:S1}
\textbf{Supplementary figures.}
\textbf{(a, b)} Related to Fig. 1 of the main text. (a) Schematic and nomenclature of amino acid pairs in offset reading frames. (b) Compatible pairs of amino acids doublets \textit{\textbf{a-b}} and \textit{\textbf{c-d}} (as defined in (a)) that can be encoded across from each other in each of the three reading frames. 
Each matrix shows in black the compatible pairs of the 400$\times$400 possible pairings.
\textbf{(c, d)} Related to Fig. 4 of the main text. Maximum compatible overlap between protein families, split by reading frame.
\textbf{(e)} Related to Fig. 6 of the main text. Distributions of hamming distance between independently sampled overlapping sequences -- of either the full coding sequence in nucleotides, or each translated gene product alone (in amino acids).
}
\end{figure*}

\clearpage

\begin{table}
    \centering
    \begin{tabular}{lll}
\toprule
Pfam ID & Description & Alignment width (aa/nt) \\
\midrule
PF00004 & AAA ATPase domain & 110/330 \\
PF00009 & GTP-binding elongation factor Tu domain & 162/486 \\
PF00011 & HSP20/alpha-crystallin small heat shock protein domain & 82/246 \\
PF00013 & KH RNA-binding domain & 58/174 \\
PF00017 & SH2 (Src Homology 2) phosphotyrosine-binding domain & 75/225 \\
PF00018 & SH3 (Src Homology 3) proline-rich peptide-binding domain & 43/129 \\
PF00041 & Fibronectin type III domain & 74/222 \\
PF00072 & Response regulator receiver domain & 108/324 \\
PF00076 & RNA recognition motif (RRM) & 59/177 \\
PF00096 & C2H2-type zinc finger domain & 23/69 \\
PF00153 & Mitochondrial carrier protein domain & 85/255 \\
PF00271 & Helicase conserved C-terminal domain & 95/285 \\
PF00397 & WW domain & 28/84 \\
PF00512 & Histidine kinase A phosphoacceptor domain & 62/186 \\
PF00595 & PDZ domain & 71/213 \\
PF02518 & Histidine kinase-like ATPase catalytic domain (GHKL) & 101/303 \\
PF07679 & Immunoglobulin I-set domain & 76/228 \\
\bottomrule
\end{tabular}
    \caption{List of PFAM alignments used in this study, as well as their width (i.e. protein length) in amino acids (aa) and nucleotides (nt).}
    \label{tab:placeholder}
\end{table}